\documentclass[english, a4paper, twoside, slac_one]{revtex4}
\usepackage{graphicx}
\usepackage{fancyhdr}
\usepackage{amsmath} 
\usepackage{bm}
\usepackage{amsxtra}
\usepackage{amssymb}
\usepackage{amsthm}
\usepackage{latexsym}
\usepackage{lscape}
\usepackage{array}

\makeatletter

\providecommand{\tabularnewline}{\\}

\makeatother
\begin{document}

\title{High-energy high-luminosity electron-ion collider eRHIC}

\author{Vladimir N. Litvinenko}
\affiliation{Brookhaven National Laboratory, Upton, NY 11973}
\affiliation{Department of Physics and Astronomy, Stony Brook University, Stony Brook 11784}

\author{Joanne Beebe-Wang}
\author{Sergey Belomestnykh}
\affiliation{Brookhaven National Laboratory, Upton, NY 11973}

\author{Ilan Ben-Zvi}
\affiliation{Brookhaven National Laboratory, Upton, NY 11973}
\affiliation{Department of Physics and Astronomy, Stony Brook University, Stony Brook 11784}

\author{Michael M. Blaskiewicz}
\author{Rama Calaga}
\author{Xiangyun Chang}
\author{Alexei Fedotov}
\author{David Gassner}
\affiliation{Brookhaven National Laboratory, Upton, NY 11973}

\author{Lee Hammons}
\affiliation{Brookhaven National Laboratory, Upton, NY 11973}
\affiliation{Department of Physics and Astronomy, Stony Brook University, Stony Brook 11784}

\author{Harald Hahn}
\affiliation{Brookhaven National Laboratory, Upton, NY 11973}

\author{Yue Hao}
\affiliation{Brookhaven National Laboratory, Upton, NY 11973}
\affiliation{Department of Physics and Astronomy, Stony Brook University, Stony Brook 11784}

\author{Ping He}
\author{William Jackson}
\author{Animesh Jain}
\author{Elliott C.Johnson}
\author{Dmitry Kayran}
\author{Jorg Kewisch}
\author{Yun Luo}
\author{George Mahler}
\author{Gary McIntyre}
\author{Wuzheng Meng}
\author{Michiko Minty}
\author{Brett Parker}
\author{Alexander Pikin}
\affiliation{Brookhaven National Laboratory, Upton, NY 11973}

\author{Eduard Pozdeyev}
\affiliation{FRIB, Michigan State University, East Lansing, MI 48824 }

\author{Vadim Ptitsyn}
\author{Triveni Rao}
\author{Thomas Roser}
\author{John Skaritka}
\author{Brian Sheehy}
\author{Steven Tepikian}
\author{Yatming Than}
\author{Dejan Trbojevic}
\affiliation{Brookhaven National Laboratory, Upton, NY 11973}

\author{Evgeni Tsentalovich}
\affiliation{MIT-Bates, Middleton, MA 01949}

\author{Nicholaos Tsoupas}
\author{Joseph Tuozzolo}
\author{Gang Wang}
\affiliation{Brookhaven National Laboratory, Upton, NY 11973}

\author{Stephen Webb}
\affiliation{Brookhaven National Laboratory, Upton, NY 11973}
\affiliation{Department of Physics and Astronomy, Stony Brook University, Stony Brook 11784}

\author{Qiong Wu}
\author{Wencan Xu}
\author{Anatoly Zelenski}
\affiliation{Brookhaven National Laboratory, Upton, NY 11973}
\maketitle

\section{Introduction }

In this paper, we describe a future electron-ion collider (EIC), based
on the existing Relativistic Heavy Ion Collider (RHIC) hadron facility,
with two intersecting superconducting rings, each 3.8 km in circumference
{[}1{]}. The replacement cost of the RHIC facility is about two billion
US dollars, and the eRHIC will fully take advantage and utilize this
investment. We plan adding a polarized 5-30 GeV electron beam to collide
with variety of species in the existing RHIC accelerator complex,
from polarized protons with a top energy of 325 GeV, to heavy fully-striped
ions with energies up to 130 GeV/u. 

Brookhaven\textquoteright{}s innovative design, (Fig. 1), is based
on one of the RHIC\textquoteright{}s hadron rings and a multi-pass
energy-recovery linac (ERL). Using the ERL as the electron accelerator
assures high luminosity in the  $10^{33}-10^{34}$ cm$^{-2}$ sec$^{-1}$
range, and for the natural staging of eRHIC, with the ERL located
inside the RHIC tunnel. The eRHIC will provide electron-hadron collisions
in up to three interaction regions. We detail the eRHIC\textquoteright{}s
performance in Section \ref{sec:sec2}.

Since first paper on eRHIC paper in 2000, its design underwent several
iterations. Initially, the main eRHIC option (the so-called ring-ring,
RR, design) was based on an electron ring, with the linac-ring (LR)
option as a backup. In 2004, we published the detailed \textquotedblleft{}eRHIC
$0^{th}$ Order Design Report\textquotedblright{} {[}2{]} including
a cost-estimate for the RR design. . After detailed studies, we found
that an LR eRHIC has about a 10-fold higher luminosity than the RR.
Since 2007, the LR, with its natural staging strategy and full transparency
for polarized electrons, became the main choice for eRHIC. In 2009,
we completed technical studies of the design and dynamics for MeRHIC
with 3-pass 4 GeV ERL. We learned much from this evaluation, completed
a bottom-up cost estimate for this \$350M machine, but then shelved
the design.

In the same year, we turned again to considering the cost-effective,
all-in-tunnel six-pass ERL for our design of the high-luminosity eRHIC
(Fig.1). In it, electrons from the polarized pre-injector will be
accelerated to their top energy by passing six times through two SRF
linacs. After colliding with the hadron beam in up to three detectors,
the e-beam will be decelerated by the same linacs and dumped. The
six-pass magnetic system with small-gap magnets {[}3{]} will be installed
from the start. We will stage the electron energy from 5 GeV to 30
GeV stepwise by increasing the lengths of the SRF linacs. We discuss
details of eRHIC\textquoteright{}s layout in Section \ref{sec:sec3}.

We considered several IR designs for eRHIC. The latest one, with a
10 mrad crossing angle and $\beta^{*}=5$ cm, takes advantage of newly
commissioned Nb$_{3}$Sn quadrupoles {[}4{]}. Section \ref{sec:sec4}
details the eRHIC lattice and the IR layout. 

The current eRHIC design focuses on electron-hadron collisions. If
justified by the EIC physics, we will add a 30 GeV polarized positron
ring with full energy injection from eRHIC ERL. This addition to the
eRHIC facility provide for positron-hadron collisions, but at a significantly
lower luminosity than those attainable in the electron-hadron mode.

\begin{figure}
\begin{tabular}{>{\centering}p{0.45\columnwidth}>{\centering}b{0.45\columnwidth}}
\includegraphics[width=0.45\columnwidth]{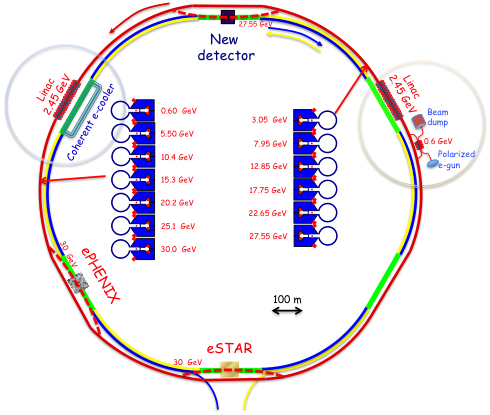}  & \includegraphics[width=0.45\columnwidth]{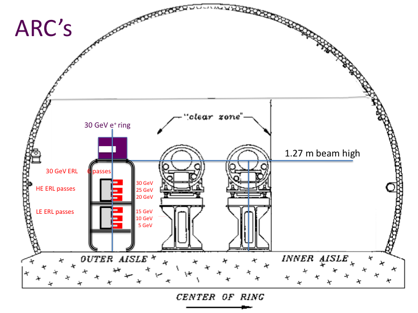}\tabularnewline
(a)  & (b)\tabularnewline
\end{tabular}

\caption{(a) Layout of the ERL based, all-in-RHIC-tunnel, 30 GeV x 325 GeV
high-energy high-luminosity eRHIC. (b) Location of six eRHIC's recirculation
arcs in RHIC tunnel.}
\end{figure}

As a novel high-luminosity EIC, eRHIC faces many technical challenges,
such as generating 50 mA of polarized electron current. eRHIC also
will employ coherent electron cooling (CeC) {[}5{]} for the hadron
beams. Staff at BNL, JLab, and MIT is pursuing vigorously an R\&D
program for resolving addressing these obstacles. . In collaboration
with Jlab, BNL plans experimentally to demonstrate CeC at the RHIC.
We discuss the structure and the status of the eRHIC R\&D in Section
\ref{sec:sec5}.

\section{Main eRHIC parameters\label{sec:sec2}}

eRHIC is designed to collide electron beams with energies from 5 to
30 GeV%
\footnote{There is no accelerator problem with using lower energy electron beams.
According to statements by EIC physicists, using electron energies
below 5 GeV would not contribute significantly to the physics goals. %
} with hadrons, viz., either with heavy ions having energies from 50
GeV to 130 GeV per nucleon or with polarized protons with energies
between 100 and 325 GeV. It means that eRHIC will cover the C.M. energy
range from 44.7 GeV to 197.5 GeV for polarized e-p, and from 31.6
GeV to 125 GeV for electron heavy-ion collisions.

Present top energy of RHIC is 250 GeV for polarized protons and 100 GeV/u for
heavy ions. We are considering a possibility of increasing these energies for up to 30\%. This increase 
is not a certainty and is a subject of dedicated studies at RHIC.

Several physics and practical considerations influenced our choice
of beam parameters for eRHIC. Some of these limitations, such as the
intensity of the hadron beam , the space charge and beam-beam tune
shift limits for hadrons, come from experimental observations at RHIC
or other hadron colliders. Some of them, for example $\beta^{*}=5$
cm for hadrons, are at the limits of current accelerator technology,
while others are derived either from practical or cost considerations. 

For example, from considering the operational costs, we limit the
electron beam\textquoteright{}s power loss for synchrotron radiation
to about 7 MW, corresponding to a 50 mA beam current at 20 GeV. Above
20 GeV, the electron beam\textquoteright{}s current will decrease
in inverse proportion to the fourth power of energy, and will be restricted
to about 10-mA at energy of 30 GeV. It means that the luminosity of
eRHIC operating with 30 GeV electrons will be a 1/5$^{th}$ of that
with 20 GeV. 

Since the ERL provides fresh electron bunch at every collision, the
electron beam can be strongly abused, i.e., it can heavily distorted
during collision. The only known effect that might cause a serious
problem is the so-called kink instability. The ways of suppressing
it within range of parameters accessible by eRHIC is well-understood
{[}6{]} and it no longer presents a problem. 

We list below some of our assumed limits and parameters: 
\begin{enumerate}
\item Bunch intensity limits:

\begin{enumerate}
\item For protons: $2\times10^{11}$; 
\item For Au ions: $1.2\times10{}^{9}$ 
\end{enumerate}
\item Electron current limits:

\begin{enumerate}
\item Polarized current: 50 mA; 
\item Unpolarized current 250 mA 
\end{enumerate}
\item 3. Minimum $\beta^{*}=5$ cm for all species 
\item Space charge tune shift limit for hadrons: $\le0.035$ 
\item Proton (ion) beam-beam parameter: $\le0.015$ 
\item Bunch length (with coherent electron cooling):

\begin{enumerate}
\item Proton: 8.3 cm at energies below 250 GeV, 4.9 cm at 325 GeV ; 
\item Au ion: 8.3 cm in all energy range 
\end{enumerate}
\item Synchrotron radiation intensity limit is defined as that of 50 mA
beam at 20 GeV 
\item Collision rep-rate $\leq50$ MHz. 
\end{enumerate}
The limitations on luminosity resulting from various considerations
are involved. The main trend is that eRHIC\textquoteright{}s luminosity
does not depend on the electron beam\textquoteright{}s energy (below
20 GeV), and reaches its maximum at the hadron beam\textquoteright{}s
highest energy. We mentioned the exception for energies of electrons
above 20 GeV. Table 1 gives the top eRHIC performance for various
species is shown in Table 1. 

\begin{table}
\caption{Projected eRHIC luminosity for various hadron beams at top energy.}

\begin{tabular}{|>{\centering}m{0.3\columnwidth}|>{\centering}m{0.1\columnwidth}|>{\centering}m{0.1\columnwidth}|>{\centering}m{0.1\columnwidth}|>{\centering}m{0.1\columnwidth}|>{\centering}m{0.1\columnwidth}|}
\hline 
 & e  & p  & $^{2}$He$^{3}$  & $^{79}$He$^{197}$  & $^{92}$He$^{238}$\tabularnewline
\hline 
\hline 
Energy, GeV  & 5-20  & 325  & 215  & 130  & 130\tabularnewline
\hline 
CM energy, GeV  &  & 80-161  & 131  & 102  & 102\tabularnewline
\hline 
Number of bunches (ions) / distance between bunches (electrons)  & 74 nsec  & 166  & 166  & 166  & 166\tabularnewline
\hline 
Bunch intensity

($\times10^{11}$ nucleons)  & 0.24  & 2  & 3  & 3  & 3.15\tabularnewline
\hline 
Bunch charge, nC  & 3.8  & 32  & 30  & 19  & 20\tabularnewline
\hline 
Beam current, mA  & 50  & 420  & 390  & 250  & 260\tabularnewline
\hline 
Normalized emittance of hadrons 95\%, mm mrad  &  & 1.2  & 1.2  & 1.2  & 1.2\tabularnewline
\hline 
Normalized emittance of electrons rms, mm mrad  &  & 5.8-23  & 7-35  & 12-57  & 12-57\tabularnewline
\hline 
Polarization, \%  & 80  & 70  & 70  & none  & none\tabularnewline
\hline 
RMS bunch length, cm  & 0.2  & 4.9  & 8.3  & 8.3  & 8.3\tabularnewline
\hline 
$\beta^{*}$, cm  & 5  & 5  & 5  & 5  & 5\tabularnewline
\hline 
Luminosity per nucleon, $10^{34}$ cm$^{-2}$s$^{-1}$  &  & 1.46  & 1.39  & 0.86  & 0.92\tabularnewline
\hline 
\end{tabular}
\end{table}

Table 2 lists the luminosity of a polarized electron-proton collision
for a set of electron and proton energies.

\begin{table}
\caption{Projected eRHIC luminosity (in $10^{33}$ cm$^{-2}$sec$^{-1}$) for
polarized electron and proton collisions.}

\begin{tabular}{|>{\centering}m{0.18\columnwidth}|>{\centering}m{0.16\columnwidth}|>{\centering}m{0.16\columnwidth}|>{\centering}m{0.16\columnwidth}|>{\centering}m{0.16\columnwidth}|}
\hline 
 & \multicolumn{4}{c|}{Protons}\tabularnewline
\hline 
Electrons  & 100 GeV  & 130 GeV  & 250 GeV  & 325 GeV\tabularnewline
\hline 
\hline 
5 GeV  & 0.62  & 1.4  & 9.7  & 15\tabularnewline
\hline 
10 GeV  & 0.62  & 1.4  & 9.7  & 15\tabularnewline
\hline 
20 GeV  & 0.62  & 1.4  & 9.7  & 15\tabularnewline
\hline 
30 GeV  & 0.62  & 0.35  & 2.4  & 3.8\tabularnewline
\hline 
\end{tabular}
\end{table}

\begin{table}
\caption{Projected eRHIC luminosity (in $10^{33}$ cm$^{-2}$sec$^{-1}$) for
polarized electrons and Au ions.}

\begin{tabular}{|>{\centering}m{0.18\columnwidth}|>{\centering}m{0.16\columnwidth}|>{\centering}m{0.16\columnwidth}|>{\centering}m{0.16\columnwidth}|>{\centering}m{0.16\columnwidth}|}
\hline 
 & \multicolumn{4}{c|}{Ions}\tabularnewline
\hline 
Electrons  & 50 GeV  & 75 GeV  & 100 GeV  & 130 GeV\tabularnewline
\hline 
\hline 
5 GeV  & 0.49  & 1.7  & 3.9  & 8.6\tabularnewline
\hline 
10 GeV  & 0.49  & 1.7  & 3.9  & 8.6\tabularnewline
\hline 
20 GeV  & 0.49  & 1.7  & 3.9  & 8.6\tabularnewline
\hline 
30 GeV  & 0.13  & 0.43  & 0.8  & 2.1\tabularnewline
\hline 
\end{tabular}
\end{table}

\begin{table}
\caption{Projected eRHIC luminosity (in $10^{33}$ cm$^{-2}$sec$^{-1}$) for
unpolarized electron and polarized proton collisions.}

\begin{tabular}{|>{\centering}m{0.18\columnwidth}|>{\centering}m{0.16\columnwidth}|>{\centering}m{0.16\columnwidth}|>{\centering}m{0.16\columnwidth}|>{\centering}m{0.16\columnwidth}|}
\hline 
 & \multicolumn{4}{c|}{Protons}\tabularnewline
\hline 
Electrons  & 100 GeV  & 130 GeV  & 250 GeV  & 325 GeV\tabularnewline
\hline 
\hline 
5 GeV  & 3.1  & 5  & 9.7  & 15\tabularnewline
\hline 
10 GeV  & 3.1  & 5  & 9.7  & 15\tabularnewline
\hline 
20 GeV  & 0.62  & 1.4  & 9.7  & 15\tabularnewline
\hline 
30 GeV  & 0.15  & 0.35  & 2.4  & 3.8\tabularnewline
\hline 
\end{tabular}
\end{table}

\begin{table}
\caption{Projected eRHIC luminosity (in $10^{33}$ cm$^{-2}$sec$^{-1}$) for
unpolarized electron and Au ion collisions.}

\begin{tabular}{|>{\centering}m{0.18\columnwidth}|>{\centering}m{0.16\columnwidth}|>{\centering}m{0.16\columnwidth}|>{\centering}m{0.16\columnwidth}|>{\centering}m{0.16\columnwidth}|}
\hline 
 & \multicolumn{4}{c|}{Ions}\tabularnewline
\hline 
Electrons  & 50 GeV  & 75 GeV  & 100 GeV  & 130 GeV\tabularnewline
\hline 
\hline 
5 GeV  & 2.5  & 8.3  & 11.4  & 18\tabularnewline
\hline 
10 GeV  & 2.5  & 8.3  & 11.4  & 18\tabularnewline
\hline 
20 GeV  & 0.49  & 1.7  & 3.9  & 8.6\tabularnewline
\hline 
30 GeV  & 0.1  & 0.34  & 0.77  & 1.7\tabularnewline
\hline 
\end{tabular}
\end{table}

Table 3 contains the same information for a polarized electron beam
colliding with Au ions, while Tables 4 and 5 provide data for the
case of unpolarized electrons.

An additional major parameter describing eRHIC\textquoteright{}s overall
performance is its expected average luminosity. Since the plans for
eRHIC are to use coherent electron cooling to control the parameters
of hadron beam, its lifetime will be affected only by scattering on
residual gas, and by burn-off in collisions with electrons. Hence,
the hadron beam\textquoteright{}s luminosity lifetime could be as
long as few days, and, in the most likely scenario, the average delivered
luminosity will be determine by the reliability of RHIC systems. Hence,
we anticipate that the average luminosity will be $\sim70\%$ of that
listed in the tables.

\section{eRHIC layout and dynamics\label{sec:sec3}}

\emph{Injector. }As shown in Fig.1, an electron gun will provide fresh
electron beams. We will employ a 50-mA polarized electron gun, based
either on single large-sized GaAs cathode {[}7{]} (Fig. 2 (a)), or
on a Gatling gun {[}8,9{]} an approach combining beams from a large
array of GaAs cathodes ( Fig.2 (b)).

\begin{figure}
\begin{tabular}{>{\centering}p{0.45\columnwidth}>{\centering}b{0.45\columnwidth}}
\includegraphics[width=0.45\columnwidth]{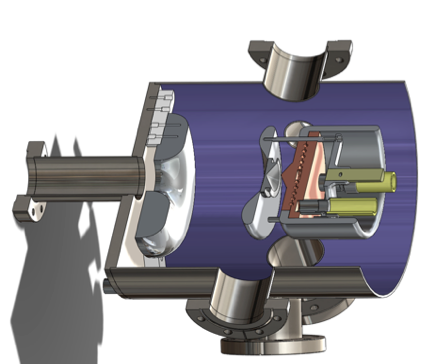}  & \includegraphics[width=0.45\columnwidth]{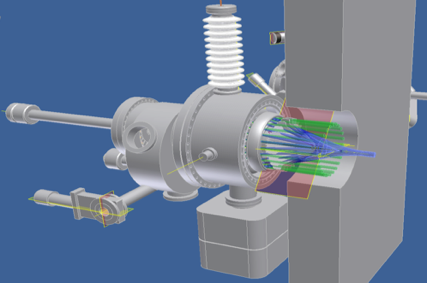}\tabularnewline
(a)  & (b)\tabularnewline
\end{tabular}

\caption{Two candidates for eRHIC polarized electron gun: (a) with a large-size
GaAs cahthode gun; (b) Gutling gun combing beams from the array of
24 GaAs cathodes. }
\end{figure}

Illuminated by circular polarized IR laser light a strained or super-lattice
GaAs cathode will produce longitudinally of highly polarized electrons.
The polarization of electrons can be as high as 85-90\%. The direction
of electron's spin can be flipped on the bunch-to-bunch basis by
changing the helicity of the laser photons.

We will utilize a dedicated un-polarized SRF electron gun, similar
to that designed for BNL\textquoteright{}s R\&D ERL {[}10{]} to generate
a significantly higher beam current (up to 250 mA CW).

Thereafter, the electrons will be accelerated in a pre-injector linac
and then will go six times around RHIC tunnel, gaining energy from
two super-conducting RF (SRF) linacs located in two of the RHIC straight
sections (see Fig. 1a, where linacs are located at 2 and 10 o'clock
straight sections). They can accommodate SRF 703 MHz linacs up to
maximum length of 201 m, which suffices for a 2.45-GeV linac operating
with a real-estate gradient of 12.45 MeV per meter, corresponding
to 20.4-MeV gain per 5-cell 703 MHz cavity.

\emph{The Main ERL.} While the magnets for the six passes around the
eRHIC will be installed from the start, the top energy of electron
beam will be raised in stages by increasing the length (and the energy
gains) of each linac in the ERL chain. At the final stage, the two
main linacs each will have energy gain of 2.45 GeV, while the injection
SRF linac will provide 0.6 GeV of energy. At all intermediate stages,
the energy gains of all linacs will be proportionally lower, i.e.,
for the10-GeV stage, the e-beam will be injected at 0.2-GeV into the
main ERL, and each main linac will provide an 0.817 GeV energy gain..

We plan to build the eRHIC\textquoteright{}s linacs from modules comprising
5-cell 703 MHz SRF cavities. Fig.3 is a 3D rendering of such modules,
and of the 5-cell cavity model with HOM dumpers. 

At their peak energy, the electrons collide with hadrons and then
their energy is recovered by the same linacs. The latter process is
assured by additional 180 degrees delay of the electrons at the top
energy, such a delay switches the acceleration to deceleration. 

\begin{figure}
\includegraphics[width=0.6\textwidth]{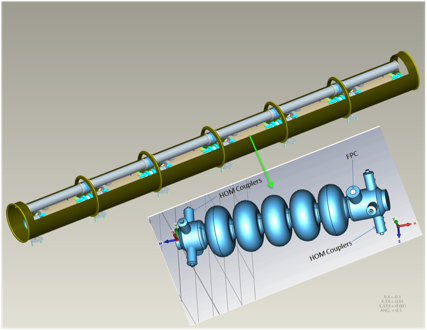}\caption{An eRHIC SRF cyro-module with 5-cell SRF cavities (insert)}
\end{figure}

\begin{figure}
\begin{tabular}{>{\centering}p{0.45\columnwidth}>{\centering}b{0.45\columnwidth}}
\multicolumn{2}{c}{\includegraphics[width=0.6\columnwidth]{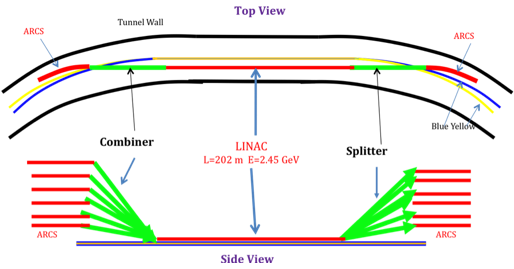}}\tabularnewline
\multicolumn{2}{c}{(a)}\tabularnewline
\includegraphics[width=0.45\columnwidth]{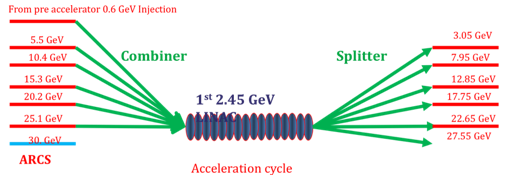}  & \includegraphics[width=0.45\columnwidth]{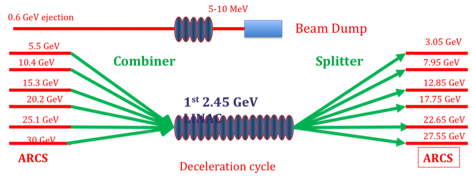}\tabularnewline
(b)  & (c)\tabularnewline
\end{tabular}

\caption{Scheme for the combiners and the splitters providing for 6 pass acceleration
and 5 pass deceleration of the electron beam in eRHIC's ERL. The
beams are separated in the vertical plane. (a) overall layout with
top and side views of the 10 o'clock RHIC straight section with eRHIC
linac; (b) action of the combiner and the splitter for accelerating
beams; (c) action of the combiner and the splitter for decelerating
beams.}
\end{figure}

\begin{figure}
\includegraphics[width=0.8\columnwidth]{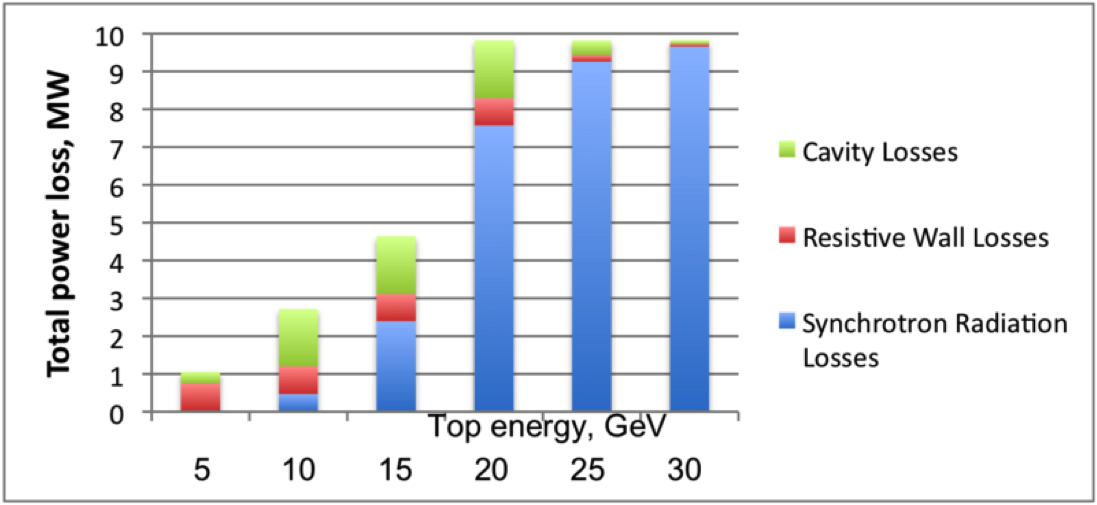}

\caption{Electron beam power loss for various top energies of eRHIC operating
with polarized electrons. Note that the losses for synchrotron radiation
are kept at the fixed level for e-beam energies above 20 GeV by reducing
the electron beam current proportionally to the forth power of the
energy. }
\end{figure}

\begin{figure}
\begin{tabular}{>{\centering}p{0.45\columnwidth}>{\centering}b{0.45\columnwidth}}
\includegraphics[width=0.45\columnwidth]{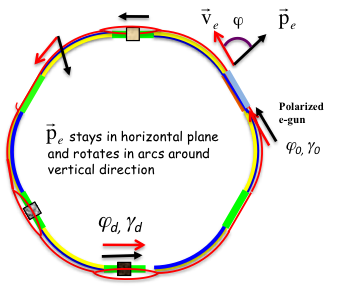}  & \includegraphics[width=0.45\columnwidth]{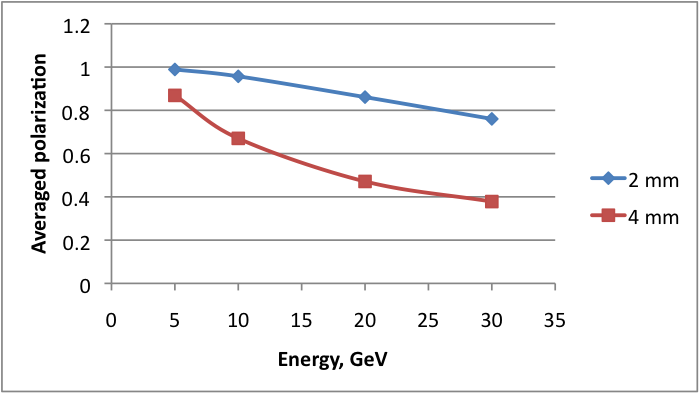}\tabularnewline
(a)  & (b)\tabularnewline
\end{tabular}

\caption{(a) Electron spin dynamics in eRHIC; (b) degree of longitudinal polarization
as function of beam energies at different bunch length.}
\end{figure}

Beams at all energies pass through the same linacs while propagating
in their individual beam-lines around the arcs. This feature is achieved
via dedicated combiners and splitters. Fig.4 depicts the arrangement
in the 10 o\textquoteright{}clock straight section; a similar system
is located in the 2 o\textquoteright{}clock section.

Except at their top energy, the accelerating and decelerating beams
share the arcs, though separated in time. For example, electron beams
at 15.3 GeV traverse the same arc between IP2 and IP10, wherein the
energy of accelerating beam increases to 17.75 GeV. It enters a 17.75
GeV arc together with the beam that just was decelerated from 20.2
GeV. In contrast, after passing through the linac, the decelerating
15.3 GeV beam passes into the 12.85 GeV arc sharing it with the beam
that was just accelerated in the same linac from 10.4 GeV. This important
ratio between accelerating and decelerating beams is maintained with
two linacs having equal energy gains. The process of the energy recovery
in SRF linacs is extremely efficient, and only about one kilowatt
of RF power per 2.45 GeV linac is needed to sustain the ERL\textquoteright{}s
operation.

The main beam-energy losses come from synchrotron radiation, resistive
losses in the walls of vacuum chambers, and HOM losses in the SRF
linacs. Figure 4 shows the values for this power loss. They must be
compensated for either by a special (second-harmonic) RF system, or
by special tuning of main linacs {[}12{]}. Additional non-compensated
beam energy results from dumping the beam at about 5 MeV; this energy
is generated by the pre-injector. 

The size of the electron beam in ERL is so small that the vertical
gap sizes in the arcs can be about a few mm; hence, using small-gap
magnets is warranted. They are important cost-saving factor for eRHIC;
we discuss the prototyping of such magnets in the section \ref{sec:sec5}.
The vacuum pipe will be made from extruded aluminum with a typical
keyhole antechamber design characteristic of modern light-sources.
In practice, the minimal vertical gap of the vacuum chamber (and,
therefore, that of the magnets) is likely to be influenced by the
tolerable wakefield effects from resistive walls and roughness. The
exact value will be determined when we theoretical and experimental
studies of these effects are completed.

\emph{Preserving polarization. }We will preserve in the ERL the high
degree of the electrons\textquoteright{} polarization originating
from the polarized electron gun {[}11{]}, and provide the desirable
direction, i.e., longitudinal, of the electron\textquoteright{}s polarization
in the interaction point (IP). The easiest (and most economical) way
of doing so is to keep the spin in the horizontal plane. In this condition,
the angle between the direction of electron\textquoteright{}s velocity
and its spin grows according a very simple equation:  
\begin{equation}
\varphi\left(\vartheta\right)=\varphi_{0}+\alpha\int_{0}^{\vartheta}\gamma\left(\theta\right)d\theta
\end{equation}
 where $\varphi_{0}$ is the initial angle at the source, $\vartheta$
is the angle of trajectory rotation in the bending magnetic field,
$\gamma=E_{e}/m_{e}c^{2}$ is the relativistic factor of electron
beam and $\alpha$ is anomalous magnetic moment of electron. By selecting
the energy of electron providing $m\pi$ total rotation angle, where
$m$ is integer between the polarized gun and the collision point
will ensure the longitudinal polarization of electrons in the IP%
\footnote{There is no need of a transverse polarization of electrons for the
physics processes of interest.%
}. As discussed, the direction of electron spin (helicity) will be
switched by reversing the helicity of the laser photons in the gun.

With six passes in ERL and layout shown in Fig. 6, the required condition
will be satisfied at IP6 for collisions at electron energies of $E_{e}=N\cdot0.07216$
GeV, where $N$ is an integer. It means that tuning energy for 0.24\%
of a top energy of 30 GeV will provide such a condition.

\emph{Arcs lattice. }The eRHIC\textquoteright{}s arc lattice has two
components, viz., that of the Blue hadron ring and the ERL lattice.
The lattice of RHIC\textquoteright{}s blue ring would be modified
significantly only in the IR straight sections. We discuss this in
next section.

\begin{figure}
\begin{tabular}{>{\centering}p{0.45\columnwidth}>{\centering}b{0.45\columnwidth}}
\includegraphics[width=0.45\columnwidth]{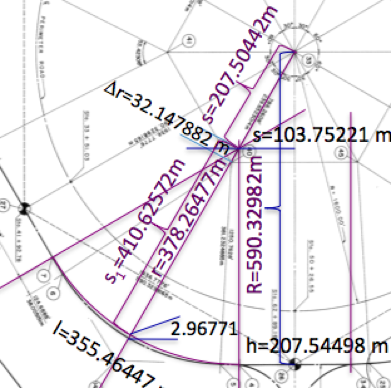}  & \includegraphics[width=0.45\columnwidth]{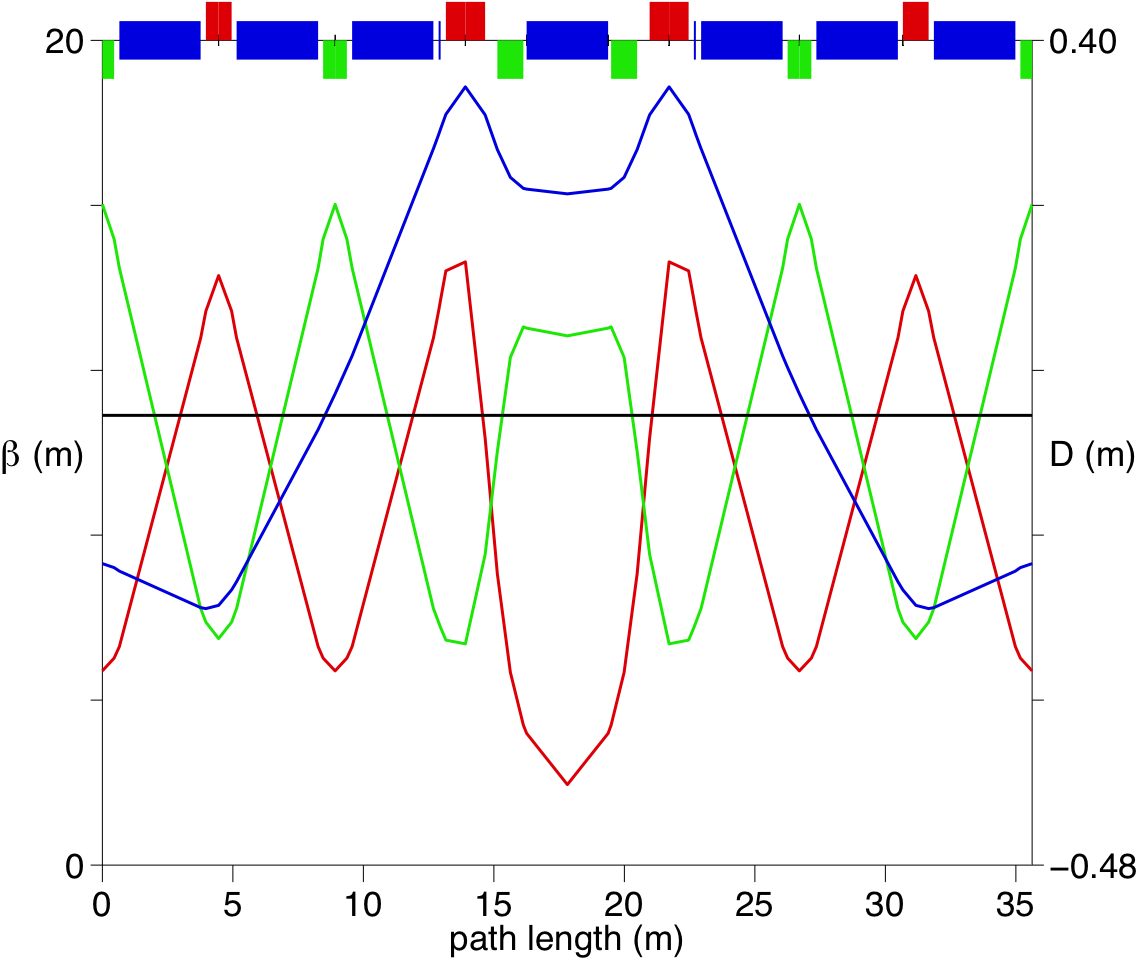}\tabularnewline
(a)  & (b)\tabularnewline
\end{tabular}

\caption{Geometry (a), and lattice functions (b) for the standard building
block. }
\end{figure}

The lattice of 6-passes for eRHIC\textquoteright{}s ERL is based on
a low-emittance near-isochronous lattice module. The concept of such
a lattice originated from the early work of Dejan Trbojevic {[}13{]}.
In addition to having an excellent filling factor, this lattice provides
for fine-tuning the R56 elements in the transport matrix, so allowing
perfect isochronism of the complete paths. Fig. 7 illustrates the
main building block of the arc lattice. Similar blocks at the both
sides of the arc lattice make it perfectly achromatic. The lattice
of the regular arcs is identical for all of them, independent of their
energy. The only differences arise from the splitters and combiners
in the SRF linac straights, as well as from the by-pass sections in
the other straights.

As evident from Fig. 4, the ERL linacs will be located inside the
RHIC rings, while ERL arcs are located outside them. This transition,
as well as other peculiarities of the RHIC tunnel\textquoteright{}s
geometry are accommodated by using two types of the same basic section
(Fig.7) with slightly different radii of curvature. 

\begin{figure}
\includegraphics[width=0.6\textwidth]{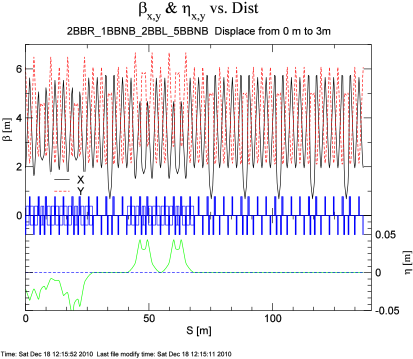}\caption{Half of the bilaterally symmetric lattice of the bypass around eRHIC
detector at 6 o\textquoteright{}clock. }
\end{figure}

\begin{figure}
\includegraphics[width=0.6\textwidth]{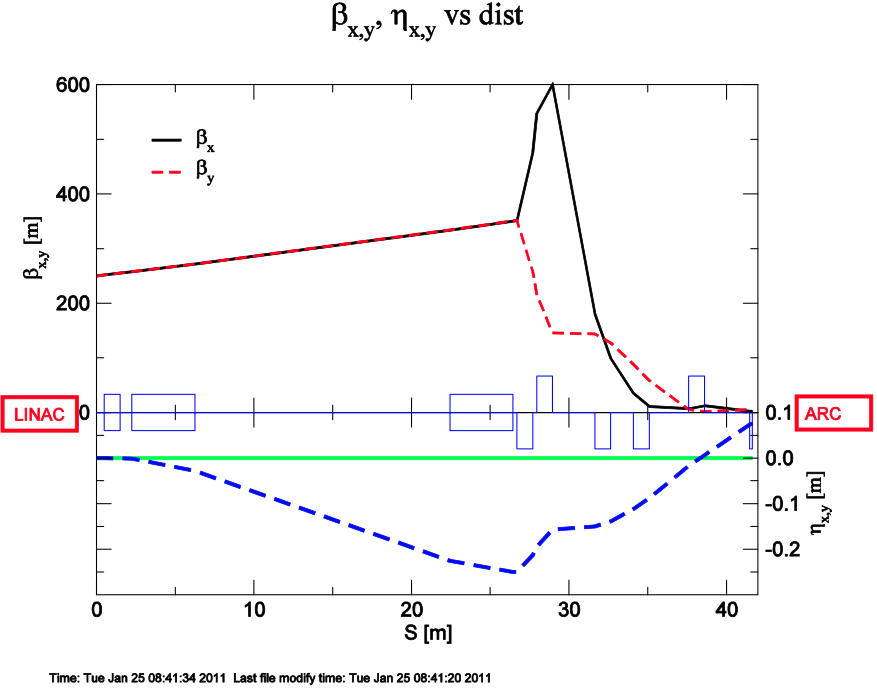}\caption{Lattice of the 30-GeV splitter matching the optical functions of the
SRF linac and the arc.}
\end{figure}

Similar basic blocks are used for straight passes and for by-passes
around the detectors. Fig. 8 shows such a design for the by-pass around
the eSTAR detector. Presently, we are considering using a linac lattice
without quadrupoles and with values of beta function of about 200
meters at its ends. Splitters and combiners serve an additional role
as matching sections between linacs and arcs. Fig. 9 shows the 30
GeV splitter matching the beta functions from the linac to the arcs. 

At present, the lattice of all six passes of eRHIC ERL is completed,
and the exact location of each ERL magnet inside the RHIC tunnel identified. 

One very important issue is finding a solution for synchronizing the
electron beam with the hadron beam circulating in RHIC at different
energies from 50- to 325-GeV/u. Being based on the ERL, eRHIC does
not suffer from standard ring-ring limitations. One elegant solution
identified is operating RHIC at energies corresponding to the hadron
beam\textquoteright{}s repetition frequency, i.e., various sub-harmonics
of the ERL RF frequency (see Fig. 10 b). The remaining tunability
of the ERL\textquoteright{}s circumference can be achieved by using
a standard eRHIC bypass in one of the free straight sections (for
example, in IP4). 

Many issues in beam dynamics for eRHIC ERL were studied and no major
deterrents were found {[}19{]}. We detailed the effects of synchrotron
radiation (both its energy spread and emittance growth), wakefields
from SRF linacs, resistive walls, and transverse beam (TBBU) stability.
We will address a few remaining questions before releasing the final
eRHIC design. A remaining one is the effect of the wakefields from
the wall\textquoteright{}s roughness on energy spread. These issues
and possible remedies are under investigation. 

\begin{figure}
\begin{tabular}{>{\centering}p{0.45\columnwidth}>{\centering}b{0.45\columnwidth}}
\includegraphics[width=0.45\columnwidth]{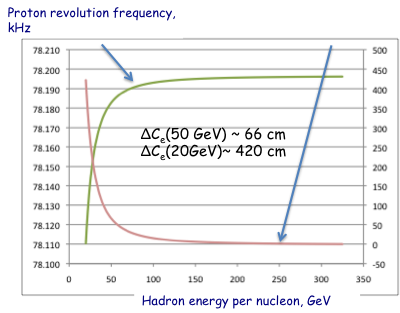}  & \includegraphics[width=0.45\columnwidth]{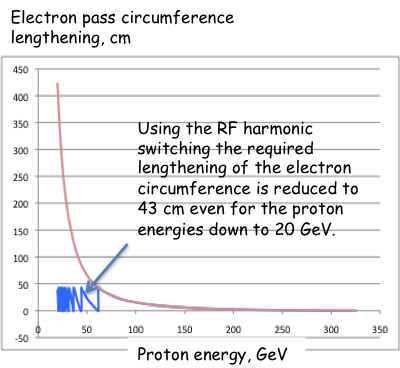}\tabularnewline
(a)  & (b)\tabularnewline
\end{tabular}

\caption{(a) Change of the revolution frequency of hadron beams in RHIC as
function of their energy; (b) Red line - the required change for the
e-beam circumference without harmonic switching (i.e. ring-ring case);
Blue \textendash{} the same curve with switching the harmonic number.}
\end{figure}

\begin{figure}
\includegraphics[width=0.6\textwidth]{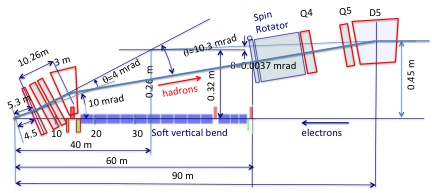}\caption{Layout of the right side of eRHIC IR from the IP to the RHIC arc.
The spin rotator is the first element of existing RHIC lattice remaining
in place in this IR design.}
\end{figure}

\section{eRHIC interaction design\label{sec:sec4}}

Current high-luminosity eRHIC IR design incorporates a 10 mrad crab-crossing
scheme; thus, hadrons traverse the detector at a 10 mrad horizontal
angle, while electrons go straight through. Fig 11 plots this scheme.
The hadron beam is focused to $\beta^{*}=5$ cm by a special triplet
wherein first magnet is a combined function magnet (1.6 m long with
2.23 T magnetic fields and a -109 T/m gradient). It has two functions;
it focuses the hadron beam while bending it 4 mrad. Two other quadrupoles
do not bend the hadron beam but serve only for focusing. Importantly,
all three magnets provide zero magnetic fields along the electron
beam\textquoteright{}s trajectory. Quadrupoles for this IR require
very high gradients, and can be built only with modern superconducting
technology {[}4,15{]}.

This configuration guaranties the absence of harmful high-energy X-ray
from the synchrotron radiation. Further, the electron beam is brought
into the collision via a 130-meter long merging system (Fig. 12).
The radiation from regular bending magnets would be absorbed. The
last 60 meters of the merging system use only soft bends: downwards
magnets have strength of 84 Gs (for 30 GeV beam) and the final part
of the bend used only 24 Gs magnetic field. Only 1.9 W of soft radiation
from the later magnets would propagate through the detector. 

\begin{figure}
\begin{tabular}{>{\centering}p{0.45\columnwidth}>{\centering}b{0.45\columnwidth}}
\includegraphics[width=0.45\columnwidth]{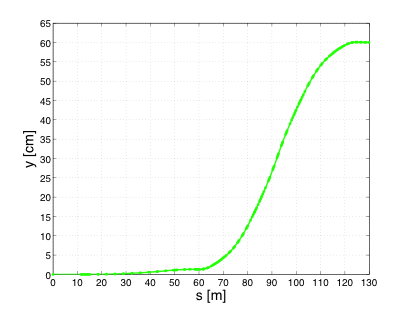}  & \includegraphics[width=0.45\columnwidth]{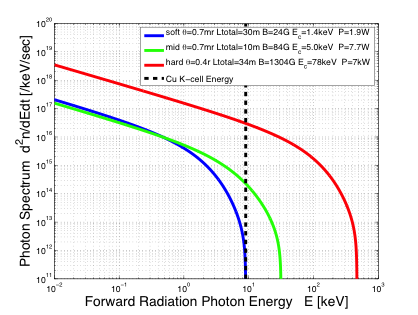}\tabularnewline
(a)  & (b)\tabularnewline
\end{tabular}

\caption{(a) Vertical trajectory of 30 GeV electron beam merging over 130 meters
into the IP. (b) Spectra of the radiation from various part of the
merger. Only 1.9 W of soft X-ray radiation will propagate through
the detector; the absorbers intercept the rest of it. }
\end{figure}

\begin{figure}
\begin{tabular}{>{\centering}p{0.45\columnwidth}>{\centering}b{0.45\columnwidth}}
\includegraphics[width=0.45\columnwidth]{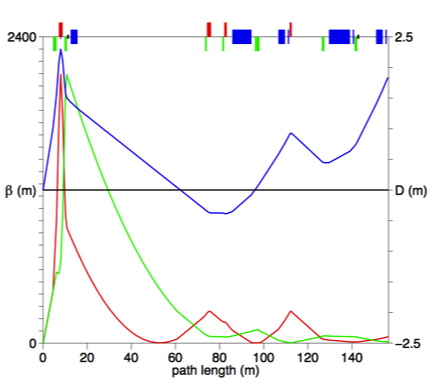}  & \includegraphics[width=0.45\columnwidth]{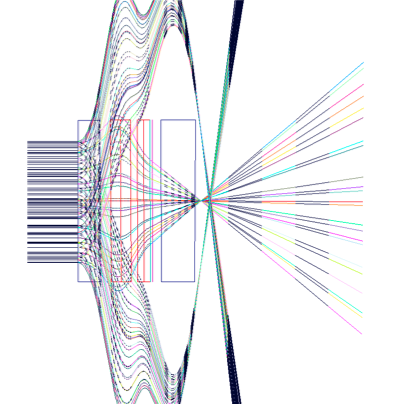}\tabularnewline
(a)  & (b)\tabularnewline
\end{tabular}

\caption{(a) Hadron beam\textquoteright{}s optics at the eRHIC IR. The 5 cm
$\beta^{*}$ is matched into the RHIC\textquoteright{}s arc lattice
that starts about 60m from the IR. (b) Tracking of hadrons with an
energy deviation of +/- 0.1\% through the first four magnets at the
IR.}
\end{figure}

One important factor in the IR design with low $\beta^{*}=5$ cm is
that the chromatism of the hadron optics in the IR should be controlled,
which is reflected in the maximum $\beta$ function of the final focusing
quadrupoles. Fig. 13a shows the designed beta and dispersion functions
for hadron beam. The values of beta function are kept under 2 km,
and the chromaticity held at the level typical for RHIC operations
with $\beta^{*}\sim1$ m. We are starting full-fledged tracking of
hadron beams in RHIC, including characterizing beam-beam effects and
all known nonlinearities of RHIC magnets: we do not anticipate any
serious chromatic effects originating from our IR design.

Furthermore, we introduced the bending field in the first quadrupole
for the hadrons thereby to separate the hadrons from the neutrons.
Physicists considering processes of interest for EIC science requested
our installing this configuration. 

Since the electrons are used only once, the optics for them is much
less constrained. Hence, it does not present any technical or scientific
challenges, and so we omit its description here. 

Finally, beam-beam effects play important role in the eRHIC\textquoteright{}s
performance. While we will control these effects on the hadron beam,
i.e., we will limit the total tune shift for hadrons to about 0.015,
the electron beam is used only once and it will be strongly disrupted
during its single collision with the hadron beam. Consequently, the
electrons are strongly focused by the hadron beam (pinch effects),
and the e-beam emittance grows by about a factor of two (disruption)
during the collision. These effects, illustrated in Fig. 14, do not
represent a serious problem, but will be carefully studied and taken
into account in designing the optics and the aperture. 

\begin{figure}
\begin{tabular}{>{\centering}p{0.45\columnwidth}>{\centering}b{0.45\columnwidth}}
\includegraphics[width=0.45\columnwidth]{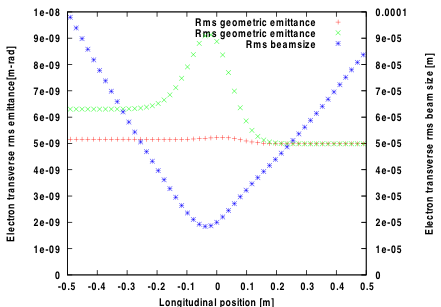}  & \includegraphics[width=0.45\columnwidth]{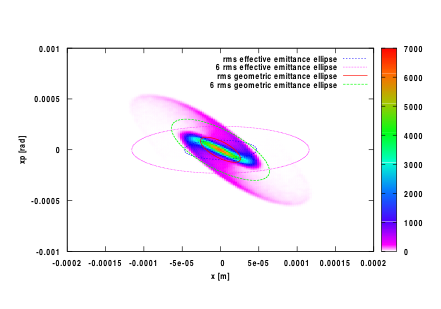}\tabularnewline
(a)  & (b)\tabularnewline
\end{tabular}

\caption{(a) The optimized e-beam envelope during collision with the hadron
beam in eRHIC; (b) distribution of electrons after colliding with
the hadron beam in eRHIC. }
\end{figure}

More details on the lattice and IR design are given in reference {[}16{]}.

\section{eRHIC R\&D\label{sec:sec5}}

The list of the needed accelerator R\&D on the eRHIC quite extensive,
ranging from the 50 mA CW polarized source to Coherent Electron Cooling
{[}5{]}. It includes designing and testing multiple aspects of SRF
ERL technology in BNL\textquoteright{}s R\&D ERL {[}18{]}. 

Coherent Electron Cooling ( Fig. 15) promises to cool both ion proton
beams to an order on magnitude smaller beams (both transversely and
longitudinally) in under a half hour. Traditional stochastic- or electron-
cooling techniques could not satisfy this demand. Being a novel unverified
technique, the CeC will be tested in a proof-of-principle experiment
at RHIC in a collaboration between scientists from BNL, JLab, and
TechX {[}17{]}. 

\begin{figure}
\includegraphics[width=0.6\textwidth]{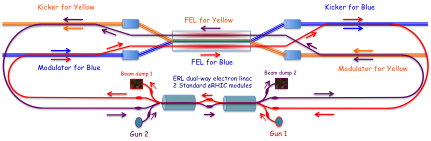}\caption{Possible layout of RHIC CeC system cooling for both the yellow and
blue beams.}
\end{figure}

Other important R\&D effort, supported by an LDRD grant, focuses on
designing and prototyping small-gap magnets and vacuum chamber for
cost-effective eRHIC arcs {[}20{]}. In addition to their energy efficiency
and inexpensiveness, small-gap magnets assure a very high gradient
as room-temperature quadrupole magnets. Fig. 16 shows two such prototypes;
they were carefully tested and their fields were mapped using high-precision
magnetic measurements. While the quality of their dipole field is
close to satisfying our requirements, the quadrupole prototype was
not manufactured to our specifications. We will continue this study,
making new prototypes using various manufacturers and techniques. 

\begin{figure}
\begin{tabular}{>{\centering}p{0.2\columnwidth}>{\centering}b{0.7\columnwidth}}
\includegraphics[width=0.2\columnwidth]{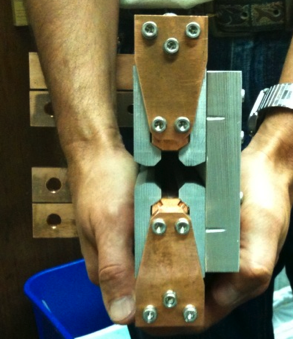}  & \includegraphics[width=0.7\columnwidth]{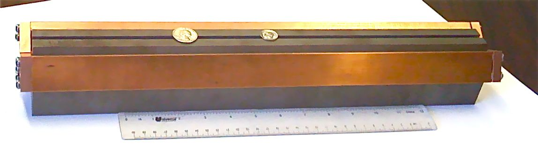}\tabularnewline
(a)  & (b)\tabularnewline
\end{tabular}

\caption{(a) A prototype of eRHIC quadrupole with 1 cm gap; (b) Assembled prototype
of eRHIC dipole magnet with 5 mm gap.}
\end{figure}

Another part of our R\&D encompasses testing the RHIC in the various
modes that will be required for the eRHIC\textquoteright{}s operation.

\section{Conclusion}

We are making steady progress in designing the high-energy, high-luminosity
electron-ion collider eRHIC and plan to continue our R\&D projects
and studies of various effects and processes. So far, we have not
encountered a problem in our proposed that we cannot resolve. Being
ERL-based collider, eRHIC offers a natural staging of the electron
beam\textquoteright{}s energy from 5-6 to 30 GeV. During this year,
we will complete our cost estimate of all eRHIC stages. 

\begin{acknowledgments}

The authors would like to acknowledge contributions and advice from
E.-C.Aschenauer, D.Bruhwiler, G.Bell, A.Cadwell, A.Deshpande, R.Ent,
W.Gurin, A.Hutton, H.Kowalsky, , G.Krafft, M.Lamont, T.W.Ludlam, R.Milner,
M.Poelker, R,Rimmer, B.Surrow, B.Schwartz, T.Ulrich, S.Vigdor, R.Venugopalan,
and W.Vogelsan.

\end{acknowledgments}

\end{document}